\documentclass[aps,prl,twocolumn,showpacs,superscriptaddress]{revtex4}
\usepackage{graphicx}
\usepackage{amsmath}
\usepackage{bm}
\usepackage{amsfonts}
\usepackage{amssymb}
\usepackage{citesort}
\begin{document}
\preprint{}
\title{Absence of low-temperature dependence of the decay of $^{7}$Be and $^{198}$Au in metallic hosts}
\author{V. Kumar}
\author{M. Hass}
\affiliation{Department of Particle Physics, Weizmann Institute of
Science, Rehovot, Israel}
\author{Y. Nir-El}
\author{G. Haquin}
\author{Z. Yungreiss}
\affiliation{Radiation Safety Division, Soreq Nuclear Research
Center, Yavne, Israel}

\date{\today}

\begin{abstract}
The electron-capture (EC) decay rate of $^{7}$Be in metallic Cu
host and the $\beta^-$-decay rate of $^{198}$Au in the host alloy
Al-Au have been measured $\it simultaneously$ at several
temperatures, ranging from 0.350 K to 293 K. No difference of the
half-life of $^{198}$Au between 12.5 K and 293 K is observed to a
precision of 0.1$\%$. By utilizing the special characteristics of
our double-source assembly, possible geometrical effects that
influence the individual rates could be eliminated. The ratio of
$^7$Be to $^{198}$Au activity thus obtained also remains constant
for this temperatures range  to the experimental precision of
$0.15\pm0.16$\%. The resulting null temperature dependence is
discussed in terms of the inadequacy of the often-used
Debye-H$\ddot{u}$ckel model for such measurements.
\end{abstract}
 \maketitle The issue
of the possible dependence of $\beta$-decay and electron-capture
(EC) rates of radioactive nuclei on the nature and temperature of
the host matrix and environment is a long standing subject
\cite{seg47,seg49}. Numerous recent experiments have claimed that
the half-life of radioactive nuclei embedded in metals would be
significantly affected by screening of the electrons in the metal
and this effect further be strengthened at low temperatures
\cite{rai05,mui06,bal06,ket06,rol06}. Several of the most recent
articles cite a longer half-life for the EC of $^{7}$Be
($0.8\pm0.2$\% \cite{wan06}) and a shorter half-life for the
$\beta^+$-decay of $^{22}$Na ($1.2\pm0.2$\% \cite{lim06}), where
these nuclides were implanted in the metals Pd and In and cooled
to $T =$ 12 K. For the $\beta^{-}$-decay of $^{198}$Au in a Au
metallic environment the half-life was observed to be longer by
$0.4\pm0.7$\% at room temperature and by $4.0\pm0.7$\% when the
metal was cooled to $T =$ 12 K, both compared to the literature
value of the half-life \cite{spi07}. The half-life of $^{7}$Be in
C$_{60}$ at $T =$ 5 K was observed to be 0.34(11)\% shorter than
that in C$_{60}$ at $T =$ 293 K and 1.5(1)\% shorter than that
inside Be metal at $T =$ 293 K \cite{oht07}. These errors are
calculated using the half-life values as provided in \cite{oht07}.
For $\alpha$-decay, the half-life of $^{210}$Po in  Cu metal at $T
=$ 12 K was measured to be shorter by $6.3\pm1.4$\% as compared to
room temperature \cite{rai07}. However, such effects could not be
confirmed by other experiments to the accuracy level of parts of a
percent \cite{zin07,sev07,cze06,rup08}.

On the theoretical side, we note that the Debye-H$\ddot{u}$ckel
screening model, used in several previous publications to explain
the apparent temperature dependence, is not applicable for a solid
(strong-coupled plasma) but rather for a weak-coupled plasma.
Furthermore, even within the Debye-H$\ddot{u}$ckel screening model
itself, the explicit temperature dependence of U$_D$ - the Coulomb
energy of the Debye cloud - of the form $1/\sqrt{T}$, is valid
only for temperatures much higher than the Fermi temperature,
$E_F$, being of the order of thousands of degrees for Cu
\cite{merm,cze06,meh07}. Below the Fermi temperature (certainly
for 293 K and lower), there should be no temperature dependence.

In light of this experimental and theoretical situation, we have
recently carried out a precision measurement of the half-life of
$^{7}$Be in different host materials, where we found a null (or
very small) change for conductors versus insulators at room
temperature \cite{nir07}. In this Letter, we report on the $\it
simultaneous$ measurement of the temperature dependence of the EC
decay rate of $^{7}$Be in metallic Cu host and of the
$\beta^-$-decay rate of $^{198}$Au in the host alloy Al-Au,
ranging from 0.350 K to 293 K. Due to the relatively short
half-life of $^{198}$Au, it is possible to determine precisely the
temperature dependence of its half-life within the time
constraints of a low-temperature experiment. Subsequently, by
utilizing the special characteristics of our double-source
assembly (see below), possible geometrical and systematic errors
that influence the individual rates could be eliminated by
examining the $\it ratio$ of $^7$Be to $^{198}$Au activities that
decay by close-energy $\gamma$-rays.
\begin{figure*}
\includegraphics[width=5.4in]{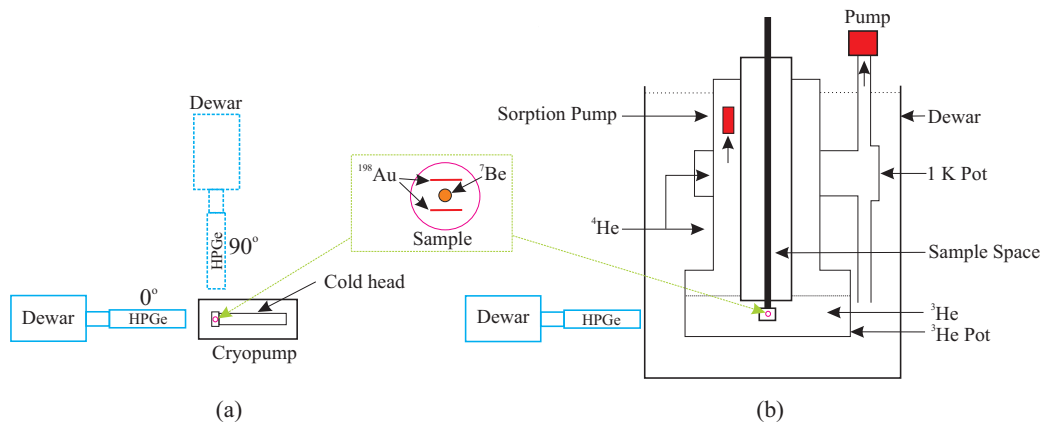}\\
\caption{(color online) (a) Schematics of the set-up for the
measurements at 12.5 K and 293 K (``experiment 1") at the
0$^{\circ}$ and 90$^{\circ}$ geometry assemblies; (b) Schematics
of the set-up for the measurement at 0.350 K and 4K (``experiment
2") at the 90$^{\circ}$ geometry assembly.} \label{fig1}
\end{figure*}

The ``double-source assembly" in the present experiment was based
on the 2 mm diameter $^{7}$Be source, prepared by direct
implantation at ISOLDE (CERN) onto a copper disk of 12 mm diameter
and 1.5 mm thickness, whose characteristics have been provided in
detail in \cite{bab03,kos03,has99}. $^{7}$Be decays by EC to the
ground and first excited state of $^{7}$Li at 477.6 keV. The
branching ratio to $^{7}$Li$^{*}$(477.6) is 10.44(4)\%
\cite{til02}. The half-life 53.353(50) d of $^{7}$Be in copper was
measured in a previous study \cite{nir07}. Adjacent to the 2 mm
spot of $^7$Be activity on the Cu disc,  a $^{198}$Au ($T_{1/2}$ =
2.6956(3) d \cite{tul05}) source was attached.  This radionuclide
disintegrates by 100\% $\beta^{-}$ emission with a 95.58(12)\%
branch of a 411.8 keV $\gamma$-ray \cite{fir96}, very close in
energy to the $\gamma$ line of $^7$Be. A 0.51 mm thick Al-Au
(0.135\% of Au) alloy wire of mass 13.7 mg was irradiated in the
nuclear reactor at the Soreq Nuclear Research Centre, Israel, to
produce $^{198}$Au by neutron activation -
$^{197}$Au(n,$\gamma$)$^{198}$Au. The 20 min irradiation in the
pneumatic transfer tube facility (``Rabbit") of the reactor
induced an activity of approximately 2.8$\times$10$^{4}$ Bq at the
end of irradiation, 1-2 days prior to the commencement of the
$\gamma$  measurements. The active Al-Au wire was glued to the
copper disk, adjacent to the $^{7}$Be spot, by a heat-conducting,
low-temperature adhesive, thus assuring identical thermal as well
as geometrical (thermally affected expansion and contraction)
properties for both $^{7}$Be and $^{198}$Au radionuclide.

Decay $\gamma$-rays were measured by a p-type coaxial HPGe
portable detector of 65\% relative detection efficiency, 2.05 keV
energy resolution (FWHM) and 70.4 peak/Compton ratio, all
specified at the 1332.5 keV $\gamma$-ray of $^{60}$Co. The
detector was enveloped by a 5.1 cm mercury cylinder placed within
a 5.1 cm thick lead shield to suppress the environmental
background.

We have used two experimental arrangements for the series of
temperature dependence measurements, one for a  $T = 293$ K to $T
= 12.5$ K comparison (``experiment 1") and one for a $T = 4$ K to
$T = 0.350$ K comparison (``experiment 2"). In ``experiment 1",
the sample was attached (see Figure \ref{fig1}(a)) to a copper
block, fixed to the cold head of a liquid-helium driven cryopump,
at a minimum temperature of $T = 12.5$ K and a typical pressure of
3$\times$10$^{-5}$ mbar. As can be seen in Fig. 1(a), the Ge
detector was placed at 0$^{\circ}$ and 90$^{\circ}$,
alternatively, to the axis of the cryopump, mounted on a massive
stand at a distance of 50 mm  from the sample for both geometries,
ensuring mechanical stability of better than 0.1 mm. The
geometrical corrections to the effective detection efficiency due
to thermal contraction of the cold head is sizably reduced when
the detector is placed at 90$^{\circ}$ (see discussion below). The
temperature was measured with a silicon diode, mounted close to
the source and was recorded continuously.

For the 0$^{\circ}$ geometry, three measurements of the $^{7}$Be
and $^{198}$Au activities at room temperature commenced
approximately 24 h after the end of irradiation in the reactor
with a typical statistical precision of the 411.8 and 477.6 keV
peaks of $^{198}$Au and $^{7}$Be of 0.04 and 0.35\%, respectively.
The cryopump was then turned on, reaching the temperature limit of
$T=12.5$ K in approximately 3 h; the counting time of each of the
next seven measurements at 12.5 K was set to about 12 h with a
typical statistical precision of 0.05 and 0.32\%. Following this
first cycle of room- and low-temperatures, the cryopump was then
turned off, with room temperature being reached in approximately 4
hours after flushing with dry nitrogen gas. A similar cycle of 6
measurements at 293 K and 8 measurements at 12.5 K was then
repeated, finally followed by 4 measurements at room-temperature
with precisions of 0.20 and 0.34\%. In the 90$^{\circ}$
measurements, there was one cycle of room- and low-temperatures
and finally, 4 measurements at room-temperature with typical
precisions of 1.08 and 0.72\% of the $^{198}$Au and $^{7}$Be
peaks, respectively. The 90$^{\circ}$ measurements resulted in
lesser precision data and were stopped when the typical
statistical uncertainty of the $^{198}$Au activity became
comparable to that of the $^7$Be activity, thus significantly
affecting the combined uncertainty on the activity ratio. The
analysis of the half-life of $^{198}$Au (see below) was based only
on the 0$^{\circ}$ measurements.

In order to probe the temperature dependence at yet lower
temperatures, we have performed a second series of experiments in
which the copper disc with the double source assembly was immersed
in a Cryofab CMSH-100 \cite{cryo} $^3$He - $^4$He cryostat (Figure
\ref{fig1}(b) - ``experiment 2"). The complete fridge includes a
$^4$He dewar which supplies liquid $^4$He to a 1 K Pot where the
$^4$He is pumped on, and a separate but thermally coupled $^3$He
system (Pot). A sorption pump (a molecular sieve) is used to pump
on the liquid $^3$He and cool it down to 0.350 K. In this way, the
sample was cooled down by being thermally coupled to the pumped-on
liquid $^3$He to reach 0.350 K and 4 K alternately (see Figure
\ref{fig1}(b)). Even though such a dewar is not well suited for
$\gamma$ detection, it nevertheless allowed probing the
temperature dependence down to 0.35 K. The Ge detector was placed
horizontally outside the dewar in the 90$^{\circ}$ geometry. Due
to the relatively large distance of the Ge detector from the
sources and due to considerable absorption of the $\gamma$
radiation in the dewar's wall, the statistical precision of the
411.8 and 477.6 keV peaks of $^{198}$Au and $^{7}$Be was not as
good as in ``experiment 1", typically 0.25 and 1.04\%,
respectively. We quote below only the $\it ratio$ of $^7$Be to
$^{198}$Au activities, a procedure that is evermore important for
this setup.

\begin{figure}[t]
\includegraphics[width=3.5in]{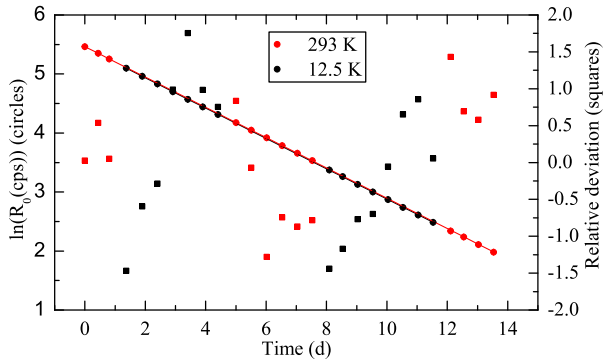}\\
\caption{(color online) Left axis-(circles): Decay curve of the
412 keV $\gamma$-ray from the $^{198}$Au in Al$-$Au. The red and 
black colors
correspond to the measured points and to the $\chi^{2}$ fitted
straight lines at 293 K and 12.5 K temperatures, respectively.
Right axis-(squares): The deviation between the measured and the
fitted count rates, divided by the corresponding individual
uncertainty. The color code is identical to that of the
decay-curve values (left-axis).} \label{fig2}
\end{figure}

The half-life ($T_{1/2}$) of $^{198}$Au at the two temperatures of
12.5 and 293 K at the 0$^{\circ}$ detector position was derived by
a weighted linear regression  of the logarithm of the net peak
count rate versus time. The decay curve of $^{198}$Au is shown in
Fig. \ref{fig2} and the calculated two straight lines fit well the
measured results. The weighted averages of the measured half-lives
of the $^{198}$Au in the host alloy Al-Au are presented in Table
\ref{table1},
\begin{table}
\caption{\label{table1} The details of the half-life determination
of $^{198}$Au and the linear regression fits at 12.5 and 293 K
temperatures.}
\begin{ruledtabular}
\begin{tabular}{cccccc}
Temperature & $n$ &  $T_{1/2}$  & ($r+1$)10$^{5}$ &
$\chi^{2}_{\nu}$ & $P_{\chi}(\chi^{2},\nu$) \\
(K) &  &  (d)&  &  & (\%) \\ \hline
293  &13 &2.6971(20) & 2.3 & 0.75 &69.0    \\
12.5  &15 &2.6976(23) & 4.9 & 1.07 &38.4    \\
\end{tabular}
\end{ruledtabular}
\end{table}
where $n$ is the number of data points, $r$ is the correlation
coefficient, $\chi^{2}_{\nu}$ is the reduced chi-square, the
number of degrees of freedom $\nu$ is equal to $n-2$, and
$P_{\chi}(\chi^{2},\nu$) is the probability that any random set of
$n$ data points would yield a value of chi-square not smaller than
$\chi^{2}$. The goodness of the linear fits is clearly shown in
Table \ref{table1}. The ratio of the half-life values at 12.5 to
293 K is calculated to be 1.0002(11) for the 0$^{\circ}$ geometry
measurement. Hence, there is a temperature null-effect of the
half-life of $^{198}$Au in the host alloy Al-Au (0.135\% of Au) at
the range 293 to 12.5 K. The weighted average of half-life of
$^{198}$Au 2.6973(15) d observed in the experiment is in agreement
within 1$\sigma$ with the published value 2.6956(3) d
\cite{tul05}. The half-life values for $^{198}$Au in agreement
with the above have been obtained also for the 90$^{\circ}$
geometries in both ``experiment 1" and ``experiment 2", albeit
with lesser precision due to the decaying source strength.

\begin{figure}
\includegraphics[width=3.55in]{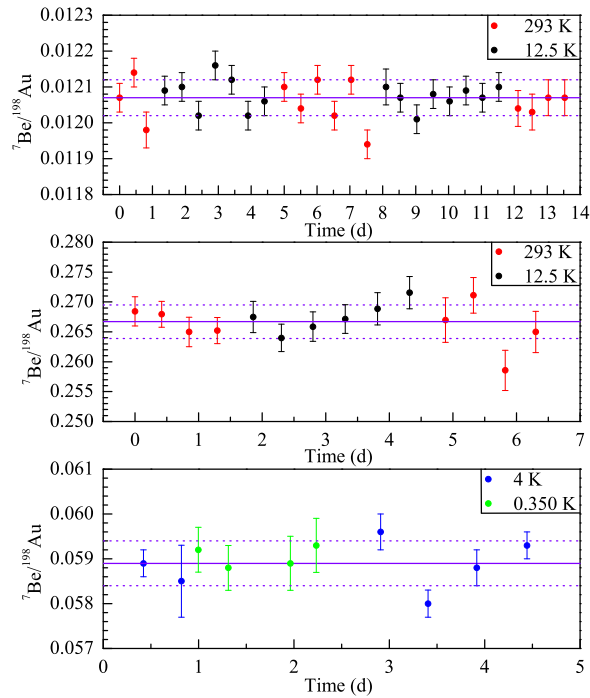}\\
\caption{(color online) Ratio of rates ($^{7}$Be/$^{198}$Au). Top:
0$^{\circ}$ geometry for 12.5 K and 293 K;
Middle:  90$^{\circ}$ geometry for 12.5 K and
293 K; Bottom:  90$^{\circ}$ geometry for 4 K
and 0.350 K (see text). The solid lines represent the weighted
averages and the broken lines correspond to an interval of
$\pm1\sigma$ standard deviation from the average - reflecting the
scattering of the individual points and possible systematic
errors.} \label{fig3}
\end{figure}

Due to the relatively low-intensity of the $^{7}$Be and the short
time span of the experiment (about 0.2 of the half-life; a
dedicated experiment for the half-life determination requires 6
months \cite{nir07}), we have not opted for a similar half-life
determination for $^{7}$Be, but have rather chosen the
\textit{normalized ratio} of $^{7}$Be to $^{198}$Au activities as
a measure of the temperature dependence of the $^7$Be EC rate. The
count rates of $^{7}$Be and $^{198}$Au were calculated for every
measured spectrum and corrected for decay using the published
half-lives 53.353 and 2.6956 d. The two reference times for the
0$^{\circ}$ and 90$^{\circ}$ measurements were the middle times of
the first room-temperature runs in each. The weighted averages of
the ratios of count rates at different temperatures are presented
in Fig. \ref{fig3} and in Table \ref{table2}.
\begin{table}[b]
\caption{\label{table2} Weighted averages of the ratio of count
rates of $^{7}$Be and $^{198}$Au, in the various temperatures and
positions of the Ge detector. First four measurements were
calculated from ``experiment 1" and last two measurements were
calculated from ``experiment 2".}
\begin{ruledtabular}
\begin{tabular}{ccc}
Temperature & Geometrical &  Ratio \\
(K) & Configuration& ($^{7}$Be/$^{198}$Au) \\\hline
293 & 0$^{\circ}$ &   0.01206(2) \\
12.5 & 0$^{\circ}$ &  0.01208(1) \\
293 & 90$^{\circ}$ & 0.2663(11) \\
12.5 &90$^{\circ}$ & 0.2672(10) \\
4    & 90$^{\circ}$ &   0.0589(3) \\
0.350 & 90$^{\circ}$ &   0.0590(3) \\
\end{tabular}
\end{ruledtabular}
\end{table}

The individual normalized count rates of $^{7}$Be and $^{198}$Au
at the 0$^{\circ}$ position for 293 K were: 2.8462(44) and
236.05(6). For 12.5 K the normalized rates were:  2.8121(29) and
232.59(20), respectively. These results demonstrate that at this
position, the count rates of both radionuclides are lower by about
($1.20\pm0.18$) to ($1.46\pm0.09$)\% at the low-temperature (12.5
K) compared to the room temperature (293 K). This finding can be
qualitatively explained by the geometrical contraction during
cooling down of the cold head of the cryopump (on which the copper
disc was mounted) that is fixed by 70 mm long stainless steel
rods. The distance between the sources and the Ge detector was 50
mm, inducing a 1.2\% decrease of the count rate, which agrees very
well with the measured decrease 1.2 to 1.5\%. This effect may have
influenced the previous measurements \cite{lim06,rai05}. The
corresponding count rates of $^{7}$Be and $^{198}$Au at the
90$^{\circ}$ position for 293 K were 0.8030(22) and 94.53(24) and
0.8092(29) and 95.57(44) for 12.5 K, respectively. This amounts to
an opposite dependence at 90$^{\circ}$, fully consistent with a
null-effect that can be anticipated since the thermal contraction
induces a negligible change in the distance between the detector
and the sources. We emphasize again that these effects cancel out
in very good approximation by the $^{7}$Be/$^{198}$Au ratios.

The half-life of $^{198}$Au in the alloy Al-Au (0.135\% of Au) was
measured and a null-effect of ($0.02\pm0.11$)\% was found at 12.5
K relative to 293 K. The count rate of $^{7}$Be in copper was
measured at these temperatures  relative to that of $^{198}$Au.
The normalized ratio of ratios (ratio $^{7}$Be/$^{198}$Au at 12.5
K to ratio $^{7}$Be/$^{198}$Au at 293 K), at the 0$^{\circ}$ and
90$^{\circ}$ are 1.0015(16) and 1.0032(58), which are equal within
the experimental uncertainty. Therefore, the half-life of $^{7}$Be
at 12.5 K is shorter by ($0.15\pm0.16$)\% than the value of 53.353
d at 293 K \cite{nir07}. The same ratios of $^7$Be to $^{198}$Au
yield 0.0590(3) and 0.0589(3) at 0.350 K and 4 K, respectively.
The ratio of ratios is therefore 1.0028(71), again a null
temperature dependence of ($0.28\pm0.71$)\% for these two
temperatures.

The presently determined decay rates for $^{7}$Be (EC)
and $^{198}$Au ($\beta^-$) for a wide temperature range of 293 K to
12.5 and 4 K to 0.35 K do not support earlier 
observations for metallic hosts and present a novel technique for
probing sub-percentage effects in the decay of, e.g., $^7$Be
and other nuclei. The results also validate the theoretical
picture above.

Note added: After submission of this manuscript, a new measurement
on the $\it T$-dependence of the half-life of $^{198}$Au has been
published \cite{Goo07}, in excellent and  full agreement with the
present null effect.

We thank Y. Shachar, S. Adam, M. Dolev and O. Zarchin of
the Weizmann Institute, and T. Riemer and the operating crew of
the nuclear reactor at Soreq for their help. We
acknowledge discussions with A. Stern and D. Salzmann. 
Supported in part by the Israel Science Foundation.

\end{document}